\documentclass[prl,aps,showpacs,twocolumn,floats]{revtex4} \input epsf
\usepackage{subfigure}
\usepackage{times}
\usepackage{graphicx,color} 
\usepackage{amsmath,amssymb}
\usepackage{natbib}
\usepackage{mciteplus}
\usepackage{bm}

\begin{document}

\title{Plastic flow in a sheared polycrystalline solid using phase field model}
\author{Santidan Biswas$^1$}
\author{Martin Grant$^2$}
\author{Indradev Samajdar$^{3}$}
\author{Arunansu Haldar$^{4}$}
\author{Anirban Sain$^{1}$}
\email{asain@phy.iitb.ac.in}
\affiliation{$^1$ Department of Physics, $^3$ Department of Metallurgical Engineering 
and Materials Science, Indian Institute of Technology, Bombay,
  Powai, Mumbai-400 076, India.\\
$^2$ Department of Physics, McGill University, Rutherford Building, Montreal H3A 2T8, Canada.\\
$^4$ Research and Development, Tata Steel, Jamshedpur 831007, India. }

\begin{abstract}
Plastic deformation in solids induced by external shear stress is of huge practical interest. 
Presence of local crystalline order in polycrystals, consisting of many grains, distinguishes  
its deformation pattern from that of amorphous materials. Despite strong anisotropy, induced by 
external stress, the plastic flow and the consequent deformation field show strong dynamical 
heterogeneity. 
The distribution $P(u)$ of particle displacements ($u$) shows three distinct regimes including a 
power law scaling regime at moderate displacements. Using a phase field simulation we show how 
polycrystals generate saddle and vortex like flow patterns, which hitherto have been termed as 
elementary plastic events in the context of amorphous materials. Interestingly, such events here 
find natural explanation in terms of the underlying dislocation dynamics. We also characterize 
the spatial distribution of the flow field using Okubo-Weiss measure.
\end{abstract}
\date{\today}

\maketitle
Solids undergo plastic deformation and subsequently flow when they are 
subjected to stresses beyond their elastic limit. Understanding 
plastic flow in polycrystalline solids has huge practical importance,
starting from applications in material processing to understanding
earthquake dynamics.
Metals are deformed routinely in various industrial applications, for
example, by rolling or pressing them into sheets. In earthquake dynamics, 
it is the shear stress generated by the movement of tectonic plates which 
causes plastic flow. Such deformations and flows under constant shear stress 
as well as constant shear strain rates have been widely studied both by lab 
experiments \cite{zapperi} and simulations \cite{chan2010}. 
Experiments on crystalline ice has established that acoustic emission which 
is a signature of dislocation avalanches in stressed solid is strongly 
intermittent and shows power law distribution of the energy released by 
such activity. Simulations using dislocation dynamics model and crystal 
phase field models have reproduced this \cite{zapperi,chan2010}. 
Simulation of amorphous solids \cite{Lemaitre1, Lemaitre2, Barrat} has 
also shown signature of spatially intermittent behavior which often 
organize into large flows in the form of vortices and saddles.

Plastic flow in a polycrystal ranges over multiple length scales starting from
the motion (glide/climb) of single dislocations at the nanometer scale 
to movement of grain boundaries at micron and larger scales. Also they are
highly anisotropic biased by the direction of the macroscopically applied
external stresses. 
We are interested in the nonequilibrium dynamics of the plastic flow which  
is generated when the solid is subjected
to external shear stress. While most studies have discussed the 
displacement field which results from a quasi-static strain \cite{Lemaitre1, 
Lemaitre2,Barrat}, we subject the solid to a constant strain rate \cite{chan2010}.

We employ phase field crystal (PFC) model to simulate a sheared polycrystalline 
solid in two dimensions. The strength of the phenomenological PFC model \cite{elder02,
stefanovic06} is that it can study dynamics of crystalline solids at the microscopic 
(atomic) length scales but diffusive time scales (much longer than vibrational 
time scales). Also here dislocations are generated spontaneously without any ad-hoc 
rules being imposed.  PFC model have been augmented (modified PFC) with an 
acceleration term in the damped Langevin equation. This reintroduces the fast 
acoustic waves into the dynamics. PFC and modified-PFC have been successful in 
reproducing important phenomenology of grain-boundary energy \cite{elder04}, 
premelting transition \cite{premelting}, dislocation glide \cite{berry06} to mention 
a few. It has also been applied to study glass transition time scales \cite{berry11} 
and liquid crystal dynamics \cite{lowen}. Modified PFC has also been derived 
\cite{elder04} from microscopic density functional theory.

\begin{figure*}[t]
\centerline{ \epsfxsize=40pc \epsfbox{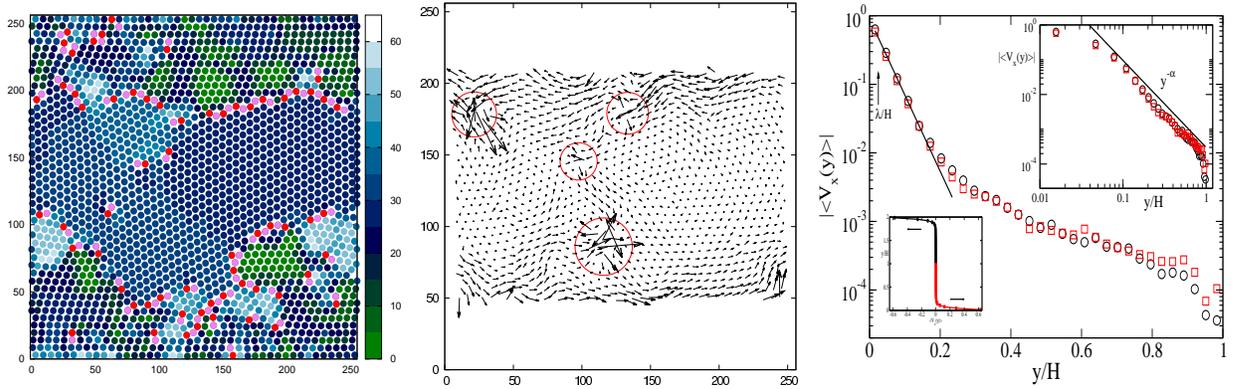} }
\caption{
(Color online) (a) Spatial map of the local crystal orientation  $\theta({\bf r})$ with a 
range $[0,\pi/3]$. Dislocations are indicated by light colors. They decorate the grain 
boundaries separating grains with high misorientation. Across grains with very low
misorientation (cells) they are absent. This is consistent with the Frank condition 
($n\propto\sin\theta$) \cite{RdSkly}, which relates the line density of dislocations 
($n$) along a grain boundary and the corresponding misorientation angle ($\theta$).
 (b) Velocity pattern of the particles in the bulk. Velocities near the
moving boundaries are too large in magnitude to be represented in the same plot. 
Vortices are visible near the top and bottom boundaries while saddles (circled) are visible 
in the bulk. We omitted the velocity pattern near the boundary region since the velocity 
scale is much higher there compared to the bulk. 
(c) Shows shows semi-log plot of $\langle v_x(y)\rangle$ as a function of $y$. To bring 
out the anti-symmetry of the average flow with respect to the mid-plane ($y=H$), we reflected 
$\langle v_x(y)\rangle$ for $y<H$ across the mid-plane and then used modulus 
(increasing $y$ is directed towards the bulk. The solid line shows the imposed drift velocity
$v(y)\sim v_0\exp(-\lambda/y)$. The upper inset shows the same data in a log-log plot which 
brings out the scaling behavior $y^{-\alpha}$ in the bulk. The lower inset shows the same data
in a regular $x-y$ plot. 
}
\vspace{.25cm}
\label{fig:test}
\end{figure*}

We confine the square shaped solid between two parallel plates, at $y=0$ and $y=2H$ 
which are moved at uniform speeds $v_0\hat x$ and $-v_0\hat x$, respectively. Along the 
$x$ direction we apply periodic boundary condition. To simulate the dynamics we use 
the modified phase field crystal (MPFC) model developed in Ref[11,17] and follow a 
shearing scheme proposed by Chan et al. \cite{chan2010}. 
The dimensionless equation of motion of the system is as follows : 

\begin{equation}
\frac{\partial^{2}{\psi}}{\partial{t}^{2}} +\beta \frac{\partial{\psi}}{\partial{t}}=
\alpha^{2} {\nabla}^2(\omega({\nabla}^2)\psi + {\psi}^3) + 
v(y)\frac{\partial{\psi}}{\partial{x}} + \zeta,
\label{eq:shearpfc}
\end{equation}

where $\psi$ is the conserved order parameter field, 
$\omega({\nabla}^2)=r+(1+{\nabla}^2)^2$ and $\zeta$ 
is the conserved noise. 
$\alpha$, $\beta$ variables control the time scale of 
the phonon modes propagating in the solid and the degree of their damping.
In this shearing scheme instead of  moving only the top and the 
bottom surfaces of the solid, a drift velocity profile, 
$v(y)\hat x=v_0\exp(-y/\lambda)$ for $0<y<H$
and $v(y)\hat x=v_0\exp(-(2H-y)/\lambda)$ for $H < y < 2H$, that decays 
exponentially away from the surface (towards the bulk) is applied 
on the solid. We hope that as long as the length scale of decay $\lambda \ll H$, 
the width of the solid, the physics in the bulk will not be affected by $\lambda$. 
We examined the effect 
of this imposed drift by measuring the average velocity along the flow 
direction, $ |\, \langle \, v_x(y) \,\rangle \, |$ as a function 
of $y$ (see inset of Fig.\ref{fig:test}-c) and inferred that there exists a 
sizable bulk portion. All our studies focus on this bulk region. 

The results presented here are from simulations on a square grid of size  
$256 \times 256$. No qualitative difference was found for a bigger grid 
($1024 \times 1024$), except that the data is much better averaged for 
the presented case due to reasonable run time. 
The parameters used are $H_x=H_y=256$, $\alpha =1$, $\beta = 1$, 
$\lambda=10$, $\psi_{0}=0.3$, $r = -0.5$, $dx = 3\pi/8$, $v_0 = 0.45$ and 
$dt=0.025$. Note that our $\lambda/H_y$ is half of that in Ref\cite{chan2010}. 
We used a pseudo-spectral scheme, combined with integrating factor
method, to solve the PDE in Eq.~\ref{eq:shearpfc}.
We used two different initial
conditions: a) a single crystal and b) a random distribution of grains of different sizes,
grown by implanting many artificial nuclei in the supercooled liquid state. Under constant 
strain rate both led to the same (statistically) nonequilibrium, steady polycrystalline 
state with relatively smaller grains near the boundaries than in the interior (see Fig.1-a). 

In PFC model, the particles are identified as the minima of the scalar field 
$\psi ({\bf r},t)$. The grains have triangular lattice structure (see Fig.\ref{fig:test}.1-a),
typical of 2D, with the local orientation field given by the angle $\theta({\bf r}) 
\in [0,\pi/3]$. The maximum misorientation between two neighboring grains can be $\pi/6$. 
Local crystal orientation at each particle was obtained from the positions of its
neighbors and keeping in mind the triangular symmetry of the crystal. 
At grain boundaries crystal orientation changes. 
Using Delaunay triangulation, the number of nearest neighbors of
each atom and the specific atoms linked to it was obtained. Dislocations were located
by finding pairs of atoms with $5$ and $7$ neighbors on the Delaunay network.

As mentioned in the introduction, collective motion in response to small shear 
strain applied at the boundaries of a solid \cite{Lemaitre1,Lemaitre2,Barrat}, 
has been studied for amorphous material. But since the applied strain there 
\cite{Lemaitre1,Lemaitre2,Barrat} was quasistatic in nature (keeping the system 
always at equilibrium or at worst metastable equilibrium), the motion was 
athermal (i.e., $T=0$). In contrast here we study the actual temporal dynamics 
in response to a constant strain rate.  
Collective motion of a many body system is often interpreted as the motion
Under a constant strain rate, as ours, the system has no time to settle down 
in the local minima consistent with the global strain. The landscape changes 
at a much faster time scale compared to the relaxational kinetics of the 
system. Thus the collective motion is nonequilibrium in nature and we focus
on the steady state features.

Several new features arise because of the polycrystalline nature of the sheared
solid. The grains do not transmit hydrodynamic stresses from one part of 
the system to the other like a fluid or even like an amorphous solids. A grain 
resists motion till the accumulated strain crosses its elastic limit and then 
either rotate with respect to its neighboring grains or break up into smaller
grains. Therefore the scale of motion is controlled by the typical grain size, 
which becomes smaller at higher strain rates. Fig.~\ref{fig:test}-b shows the 
detailed velocity map of the particles in the bulk. 
Despite the strong bias along $\hat x$ (the shear direction) the flow field 
shows significant motion along $\hat y$, giving rise to characteristic vortical 
and saddle type of motion as shown in Fig.~\ref{fig:test}-b. Origin of such kind
of motion will be discussed later. The inset of 
Fig.~\ref{fig:test}-c shows $\langle v_x(y)\rangle$ as a function of $y$, which 
shows an emergent velocity profile in the bulk differing from the imposed drift
velocity.  
Despite the shear induced anisotropy the flow is highly heterogeneous 
even within the bulk: very slow in the interior of the grains and fast at the
grain boundaries. Since flows are driven by local stresses, high shear stresses 
are expected at the grain boundaries which we will discuss next. The plastic flow 
also results in interesting displacement patterns which will be discussed later.

\begin{figure*}[t]
\centerline{\epsfxsize=12pc \epsfbox{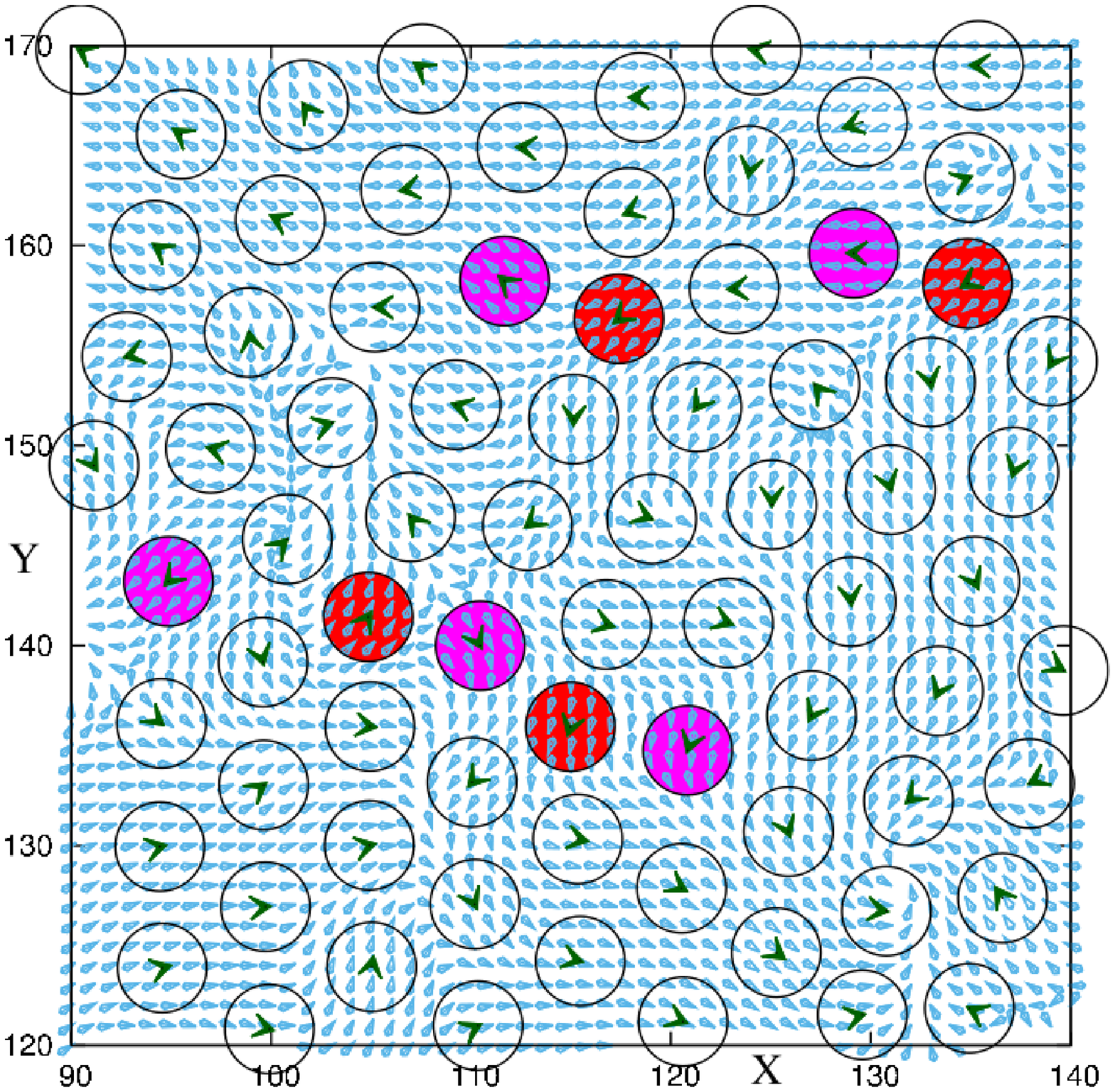} \hfill\hspace*{-4.5cm} 
\epsfxsize=18pc \epsfbox{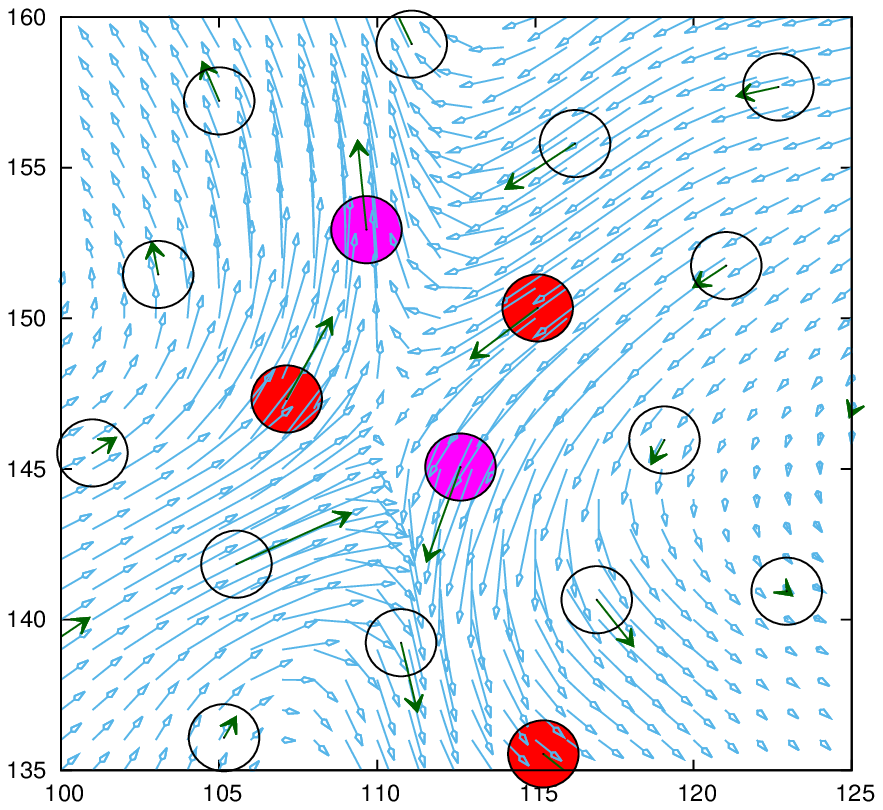}\hfill\hspace*{-4.5cm} 
\epsfxsize=11.5pc \epsfbox{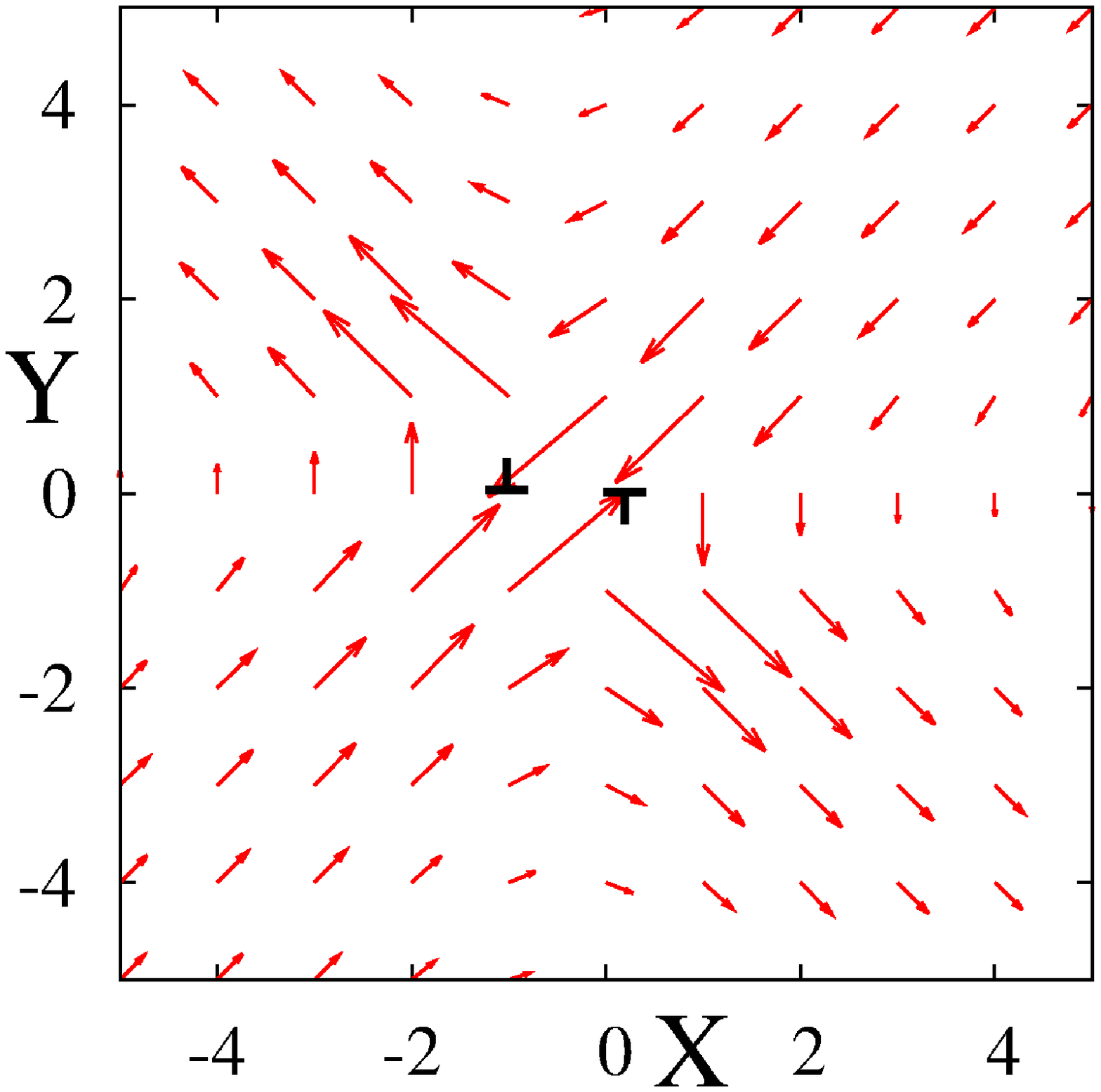 } }
\caption{(a) and (b) focus on a small region of the lattice. The circles
represent the atoms and the dark arrows on the circles represent their velocity
vectors while the light arrows represent the interpolated velocity field.  
The filled circles are the atoms with five or seven neighbors, indicating
an edge dislocation. (a) shows initiation of a saddle as two oppositely 
charged dislocations approach. (b) Time development of the saddle as the 
dislocations get close to form a quadrupole. Note that 'b' has higher 
resolution than 'c'.
(c) The superposed displacement field of two edge dislocations 
(with opposite Burgers vector) calculated using Eq.~(\ref {eq:dipole}). 
The parameters used are $b=1$ and $\nu = 0.1$. The dislocation positions 
are $(0,0)$ and $(-1,0)$.}
\label{fig:dipole}
\end{figure*}

Large scale molecular dynamics (MD) simulations on sheared amorphous material 
\cite{Lemaitre1, Lemaitre2, Barrat} have demonstrated that plastic displacement 
patterns organize into large scale vortices when the externally imposed, global, 
shear strain is changed quasi-statically. Ref.~\cite{Lemaitre2} has shown that an 
elementary, irreversible, plastic event is a quadrupolar displacement pattern 
which has long range elastic effect. Picard et al \cite{Picard}, studied the 
effect of a localized plastic event in a sheared 2D viscoelastic medium. They 
analytically showed how, through elastic interaction, it induces a long range 
strain field which turns out to of quadrupolar nature. The event was assumed to 
be a localized plastic strain $\epsilon _{ij}\delta({\bf r})$ corresponding to a 
pure shear deformation, i.e., only the non-diagonal elements of $\epsilon _{ij}$ 
were non-zero and equal. 
Generation of such a plastic strain matrix is at best plausible given 
the material is being globally sheared through its external boundaries. 
The MD simulations of Maloney and Lemaitre \cite{Lemaitre1,Lemaitre2} 
on 2D sheared amorphous materials could nail down such local plastic events to 
characteristic motion of particles near a saddle with one stable 
and another unstable axis, and  particles move towards and away, respectively, 
from the saddle along these axes. 
But it is still unclear how such singular points 
are created in the interior of the system due to global shear stress 
applied at its distant boundaries. 

\begin{figure*}[t]
\centerline{ \epsfxsize=13.75pc \epsfbox{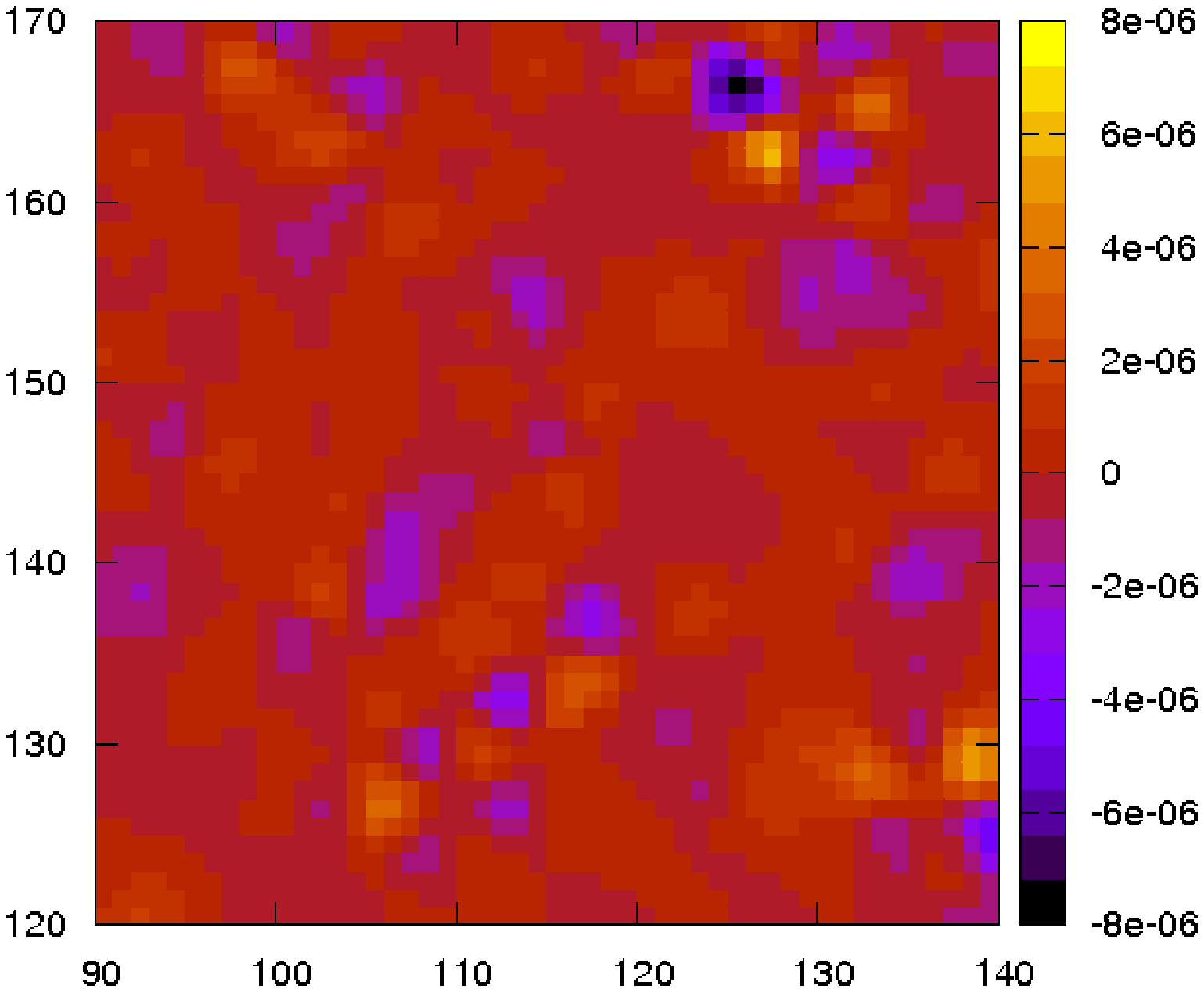} \hfill\hspace*{-2.5cm}
\epsfxsize=14pc \epsfbox{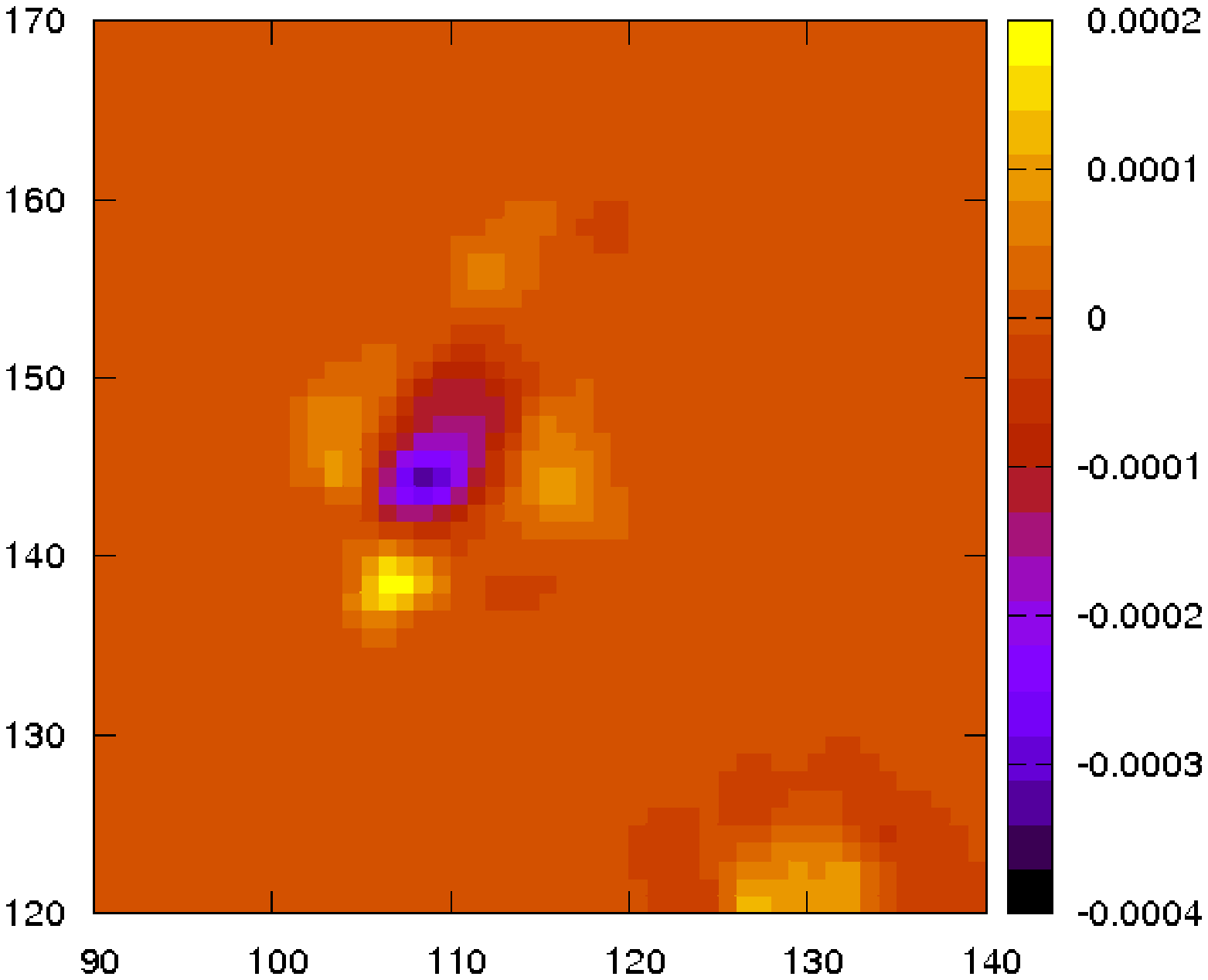}\hfill\hspace*{-2.5cm}
\epsfxsize=13.25pc \epsfbox{okubows.eps} }
\vspace{.2cm}
\caption[okubo-weiss]{(a) Okubo-Weiss field $\lambda(x,y)$ in the bulk, corresponding to the velocity
fields a and b of Fig.\ref{fig:dipole}. (b) clearly shows the prominent saddle (violet) and the 
surrounding vorticity field (yellow). Note that, in (a) also many saddles (violet) are visible 
but their intensity is two order of magnitude weaker than that in (b). 
(c) Log-log plot of the distribution of the Okubo-Weiss field. We show separate PDFs' for the 
positive (vorticity) and negative (extensional) values of $\lambda$, indicated by $\lambda_+$ 
and $\lambda_-$ respectively ($\lambda_-=|\lambda|$ when $\lambda<0$).  The PDFs' are nearly 
symmetric and has power law regimes. 
}
\label{fig:okuboweiss}
\end{figure*}

We identify a microscopic mechanism which can dynamically give rise to 
quadrupolar patterns, essentially a saddle, in the displacement field of 
the particles.  We show that such a pattern results from sideways 
approach of two oppositely charged edge dislocations towards each other. 
The sequence in Fig.~\ref{fig:dipole} clearly shows time development of the 
displacement field leading towards a saddle as the dislocations approach 
each other. The saddle fades away after the dislocations annihilate (not shown 
here).  Such long range quadrupolar patterns have been
reported for noncrystalline (amorphous) material \cite{Lemaitre2} also, 
but there being no dislocations in noncrystalline material microscopic origin of 
such a quadrupolar pattern remained unexplained. Quantitatively we know that the 
displacement field of an edge dislocation generates a displacement dipole 
where the positive and negative lobes are oriented along the axis connecting 
the atoms with coordination numbers $5$ and $7$. Such a $5-7$ pair is like 
a charge dipole and sideways arrangement of two such pairs form a quadrupole.
Essentially these two dislocations have opposite burgers vectors $\vec b$ 
and $-\vec b$ along one of the symmetry axes of the crystal. In comparison, 
dislocations with the same burgers vector can line up in an ``..5-7-5-7-5-7.." 
arrangement to form a dislocation wall (see Fig.~\ref{fig:test}-a) which is 
rather stable. These walls are equivalent to high angle grain 
boundaries (Fig.~\ref{fig:test}-a). The quadrupolar structure 
discussed above can be quantitatively established by superimposing the elastic 
displacement fields of two dislocations located close by. Fig.~\ref{fig:dipole} 
shows the resultant field from two dislocations located at ${\vec r}=(0,0)$ 
and $(-1,0)$, where displacement field $\vec u({\bf r})$ 
due to a  dislocation at the origin is given by \cite{Timoshenko}

\begin{figure}[!htb]
\scalebox{0.4}{\includegraphics[trim=0cm 0cm 0cm 0cm, clip=true]{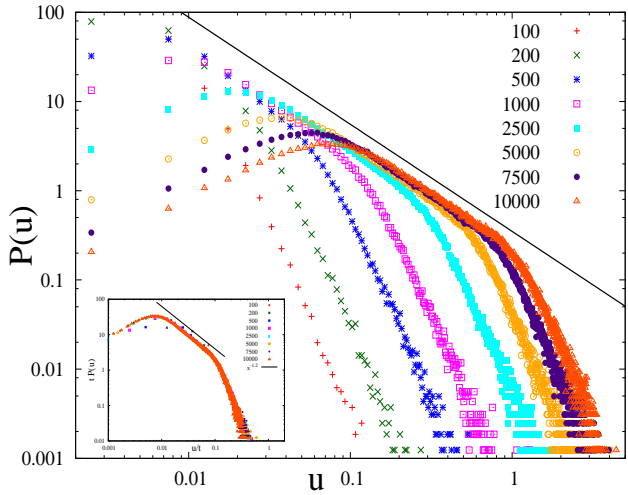}}
\caption{ Distribution of displacement magnitudes of all the particles, collected after different 
time intervals $t$ (shown in arbitrary units in the figure). The inset shows data collapse of 
the plots shown in  after rescaling $u$ by $t$, and $P(u)$ appropriately. This collapse is possibly
a consequence of the fact that the average $x-$displacement $\langle u_x\rangle=\langle v_x\rangle t$ 
dominates the behavior of $u$. But the scaling behavior cannot be explained by this. 
The scaling exponent is $-1.2$ and the plots for large $t$ only show significant scaling regimes.}
\label{fig:veldistribution}
\end{figure}

\begin{eqnarray}
u_x&=&\frac{b}{2\pi}\Big[ \tan ^{-1}\frac{y}{x} + \frac{xy}{2(1-\nu)(x^2+y^2)}\Big],\nonumber\\
u_y&=&-\frac{b}{2\pi}\Big[ \frac{1-2\nu}{4(1-\nu)}\ln(x^2+y^2) +
\frac{x^2-y^2}{4(1-\nu)(x^2+y^2)}\Big],
\label{eq:dipole}
\end{eqnarray}
where $\nu$ is the Poisson ratio and $\tan ^{-1}\frac{y}{x}\in[0,2\pi]$. It is 
worth mentioning that we have found saddle structures in the displacement profile even in places where 
there are no dislocations. Thus the above mechanism cannot be the only 
reason for saddles.

In order to study the spatial distribution of vorticity and saddles (centres) in our 2D plastic 
flow  we employ a technique borrowed from fluid turbulence \cite{Weiss,Perlekar}. 
For 2D inviscid, incompressible flows the Okubo–Weiss parameter is defined as 
$\lambda = \det (\partial _i v_j)$. This is an invariant of the flow and
can be recast as $\lambda = \omega ^2 - \epsilon ^2$, where $\vec \omega =\vec\nabla\times \vec v$ 
is the vorticity vector and $\epsilon^2= \sum_{i,j} \epsilon_{ij}^2$, where $\epsilon_{ij}=
(\partial_iu_j+\partial_ju_i)/\sqrt{2}$ is the strain tensor. Even in viscous flows $\lambda$ turns
out to be an useful measure and regions with vortices have $\lambda>0$, while the strain dominated
regions have $\lambda<0$. Note that a saddle correspond to stretching in one direction and compression
in the orthogonal direction, i.e., it generates a strain dominated region. 

Our system is not strictly incompressible, so we computed the incompressibility parameter 
$\kappa= \langle (\vec \nabla . \vec v)^2\rangle/ \langle (\vec \nabla \vec v)^2\rangle$, 
where the denominator is essentially $\sum_{i,j}(\partial _i v_j)^2$. For an incompressible 
fluid the numerator is zero, while our system yields a value of $\kappa<0.2$ (averaged over 
many configurations), which is small enough for an incompressibility approximation to be valid. 
Also in terms of total particle numbers, the fluctuation is less than $1\%$ (less than $20$ in 
$2000$). For computing $\kappa$ we interpolated the particle velocities 
onto a square grid. In Fig\ref{fig:okuboweiss}-a,b we plot the Okubo-Weiss field corresponding to 
the velocity fields (a and b) of Fig.\ref{fig:dipole}. We also compute the distribution function
of $\lambda$, shown in Fig.\ref{fig:okuboweiss}-c.

Finally we report intriguing power laws in the PDF of the particle displacements 
$|\vec u_j|$, in the bulk (excluding the boundary region where $\langle v_x(y)\rangle$ is
exponential). Here $j$ is the particle index. It turns out 
that although velocity of the particles, $\vec v_j$, are quite random, 
the displacements $\vec u_j$ after large time intervals show characteristic patterns, 
around the plastic events (figure not shown here). 
The PDF of $u \equiv |\vec u_j|$  is shown in Fig.~\ref{fig:veldistribution},
which, at large time intervals $t$, shows two clear power law regimes.
Rescaling $u$ with $t$ (and also $P(u)$ appropriately) the PDFs' collapse nicely 
(inset of Fig.4), although the PDFs' for short $t$ do not have any power law regime.

The different scaling regimes of $P(u)$, namely, $\propto u$ and $\propto u^{-1.2}$,  
reflect distinct kind of particle motion in the sheared polycrystal. $P(u)$ 
is the fraction of particles undergoing particular type of motion and is therefore 
approximately proportional to the area fraction occupied by these particles in
a typical velocity map like Fig.2b. 
The displacements are small at the core of the large grains where
motion is vortical. Assuming a slow rotational speed $\omega_0$, the displacement
$u$, for $\omega_0 t \ll 1$, is $u\sim \omega_0 rt$, where the radius $r$ 
is measured with respect to the center of the grain. Thus $P(u)du\propto dA=2\pi r dr$ 
and using $u\sim \omega_0 rt$, we get $P(u)\sim u \frac{2\pi}{(\omega_0 t)^2}$. Consequently
a time independent collapse occurs in the $P(u)t$ versus $u/t$ plot (inset of Fig.4). 
Larger displacements (the $u^{-1.2}$ regime) are dominated by $\langle v_x \rangle$ and 
here approximately $\langle v_x(y) \rangle\sim y^{-3}$ (log-log plot of 
$\langle v_x(y)\rangle$ versus $y$ not shown here). Now in this case
$P(u)du\propto dA=dy.L$ and using $u\sim y^{-3}t$, we get $tP(u)\sim \frac{1}{3} 
(\frac{u}{t})^{-4/3}$; again a $t$ independent collapse. Although our $P(u)$ is 
restricted to the bulk region, any remnant effect from the boundary region, 
where $\langle v_x(y) \rangle$ is exponential, would contribute a $u^{-1}$ scaling.  
The observed $u^{-1.2}$ scaling is possibly a mixed effect.

In summary, we have shown that the plastic flow in sheared polycrystals
show strong dynamical  heterogeneity which manifests as three distinct regimes
in the displacement distribution of the particles. Further, the elementary
plastic events of the flow field can be explained in terms of the underlying dislocation
dynamics. That bridges two seemingly disparate descriptions, namely continuum and
discrete, of sheared solids.

\bibliographystyle{h-physrev}
\bibliography{paperPlasticFlow}{}

\begin{thebibliography}{10}

\bibitem{zapperi}
M.-C. Miguel, A.~Vespignani, S.~Zapperi, J.~Weiss, and J.-R. Grasso,
\newblock Nature {\bf 410}, 667 (2001).

\bibitem{chan2010}
P.~Y. Chan, G.~Tsekenis, J.~Dantzig, K.~A. Dahmen, and N.~Goldenfeld,
\newblock Phys. Rev. Lett. {\bf 105}, 015502 (2010).

\bibitem{Lemaitre1}
C.~Maloney and A.~Lemaitre,
\newblock Phys. Rev. Lett. {\bf 93}, 016001 (2004).

\bibitem{Lemaitre2}
C.~Maloney and A.~Lemaitre,
\newblock Phys. Rev. Lett. {\bf 93}, 195501 (2004).

\bibitem{Barrat}
A.~Tanguy, J.~P. Wittmer, F.~Leonforte, and J.-L. Barrat,
\newblock Phys. Rev. B {\bf 66}, 174205 (2002).

\bibitem{elder02}
K.~R. Elder, M.~Katakowski, M.~Haataja, and M.~Grant,
\newblock Phys. Rev. Lett. {\bf 88}, 245701 (2002).

\bibitem{stefanovic06}
P.~Stefanovic, M.~Haataja, and N.~Provatas,
\newblock Phys. Rev. Lett. {\bf 96}, 225504 (2006).

\bibitem{elder04}
K.~R. Elder and M.~Grant,
\newblock Phys. Rev. E {\bf 70}, 051605 (2004).

\bibitem{premelting}
J.~Berry, K.~R. Elder, and M.~Grant,
\newblock Phys. Rev. B {\bf 77}, 224114 (2008).

\bibitem{berry06}
J.~Berry, M.~Grant, and K.~R. Elder,
\newblock Phys. Rev. E {\bf 73}, 031609 (2006).

\bibitem{berry11}
J.~Berry and M.~Grant,
\newblock Phys. Rev. Lett. {\bf 106}, 175702 (2011).

\bibitem{lowen}
R.~Wittkowski, H.~L\"owen, and H.~R. Brand,
\newblock Phys. Rev. E {\bf 82}, 031708 (2010).

\bibitem{RdSkly}
W.~T. Read and W.~Shockley,
\newblock Phys. Rev. {\bf 78}, 275 (1950).

\bibitem{Picard}
G.~Picard, A.~Ajdari, F.~Lequeux, and L.~Bocquet,
\newblock Eur. Phys. J. E {\bf 15}, 371 (2004).

\bibitem{Timoshenko}
S.~Timoshenko,
\newblock {\em Theory of Elasticity} ({McGraw-Hill}, 1951).

\bibitem{Weiss}
J.~Weiss,
\newblock Physica D {\bf 48}, 273 (1992).

\bibitem{Perlekar}
P.~Perlekar and R.~Pandit,
\newblock New J. Phys. {\bf 11}, 073003 (2009).

\end{thebibliography}
\end{document}